\documentclass[]{iau}
\usepackage{graphicx}

\newcommand{\Msun}{$M_\odot$}
\newcommand{\izwnw}{I\,Zw\,18-NW}

\shorttitle{Properties of Primitive Galaxies}

\title{Properties of Primitive Galaxies}
\author[Heap et al.]
{Sara R. Heap$^1$,
I. Hubeny$^2$, 
J.-C. Bouret$^3$, 
T. Lanz$^4$, 
\and \\ J. Brinchmann$^5$}
\affiliation{$^1$Emerita scientist, NASA Goddard Space Flight Center, Greenbelt, MD,\\[\affilskip]
$^2$Steward Observatory, University of Arizona, Tucson, AZ,\\[\affilskip]
$^3$Aix-Marseille Univ. CNRS, CNES, LAM, Marseille, France,\\[\affilskip]
$^4$Observatoire de C\^{o}te d'Azur, Nice, France,\\[\affilskip]
$^5$University of Porto, Porto, Portugal}

\pubyear{2022}
\volume{361} 
\jname{Massive Stars Near \& Far}
\editors{}

\begin{document}

\maketitle

\begin{abstract}

We report on a study of 9 nearby star-forming, very low-metallicity  galaxies observed by Hubble's COS far-UV spectrograph that can serve as templates of high-z galaxies to be observed by JWST. We find that the nebular spectra of these primitive galaxies show evidence of irradiation by X-ray emitters. Following Thuan et al. (2004), we identify the sources of X-ray emission as massive X-ray binaries containing a massive accreting stellar black hole. We further find that the lower the metallicity, the higher the probability of strong X-irradiation.  Following Heger et al. (2003), we suggest that these accreting black holes are produced by direct collapse of stars having  initial masses greater than $\sim50\, M_\odot$. Our models of young star clusters with an embedded stellar black hole produce effects on the surrounding gaseous medium that are consistent with the observed spectra. We conclude that primitive galaxies are qualitatively different from more metal-rich galaxies in showing evidence of hard radiation that can best be explained by the presence of one or more embedded stellar black holes. 
   
\end{abstract}

\section{Introduction}

We present a study of far-UV spectra of nearby primitive galaxies that could give insights into processes operating in the high-z galaxies thought to be responsible for the re-ionization of the universe.  By ``primitive galaxies,'' we mean galaxies having a low stellar mass, $\log (M_\ast/M_\odot) \leq 8$ and low gas metallicity, $\log O/H+12 \leq 8$ and an even lower iron abundance, whether  they are local or at high redshift. 

Our main goal here is to gain insight into the role of metallicity in
influencing the physical properties, evolutionary paths, and destinies
of massive stars and through them, with the help of models, an
understanding of  low-metallicity galaxies both near and far.
One approach is to examine the high-velocity winds of hot, massive
stars. Another approach is to test the importance of X-ray radiation
in very low metallicity galaxies, which
was first pointed out by  Thuan et al. (2004) who found  that the three most metal-deficient galaxies are ultra-luminous X-ray emitters (ULX). A ULX is defined by having an X-ray luminosity, $L_x>1\times 10^{39}$ erg s${}^{-1}$. They identified these X-ray sources as massive X-ray binaries (MXRB's), each  containing a massive, accreting stellar black hole. The optical spectral region also shows evidence of hard radiation. Shirazi \& Brinchmann (2012) found that spectra of primitive galaxies observed by the SDSS show \textit{nebular} He~II $\lambda 4686$ in emission, the result of ionization to He~III and subsequent recombination. Recently, spectroscopic surveys by Izotov et al. (2021) found six compact star-forming galaxies with 12+log O/H=7.46-7.88 all showing evidence of hard radiation in the form of high-ionization emission lines such as [Ne~V].

\section{Observations of primitive galaxies}

The far-UV spectral region is the most informative region for studying stellar winds of primitive stars  as it has resonance lines of abundant elements in several ionization stages. There is empirical evidence from far-UV spectra of hot, massive stars in the galaxy and Magellanic Clouds that the strength of the stellar wind increases with increasing metallicity as $Z^{0.7}$ (Vink et al. 2001), but this relation has not been tested below $\log Z\sim 8.0$. We have made predictions of the far-UV spectra of such primitive stars based on the grid of non-LTE photospheric models by Lanz \& Hubeny (2003, 2007) and  unified spectra (photosphere + wind) calculated by Bouret using the CMFGEN code (Hillier \& Miller, 1998). Figure 1 shows predicted spectra of massive stars demonstrating that even at $\log Z/Z_\odot=-2$, the N~V doublet and C~IV doublet are still detectable stellar-wind features. When compared to the spectrum of the most metal-deficient galaxy in our sample, \izwnw, there is qualitative agreement between theory and observation, given that we are comparing the spectra of a galaxy to that of a single star.

\begin{figure}[h]
\begin{center}
\includegraphics[width=5.5in]{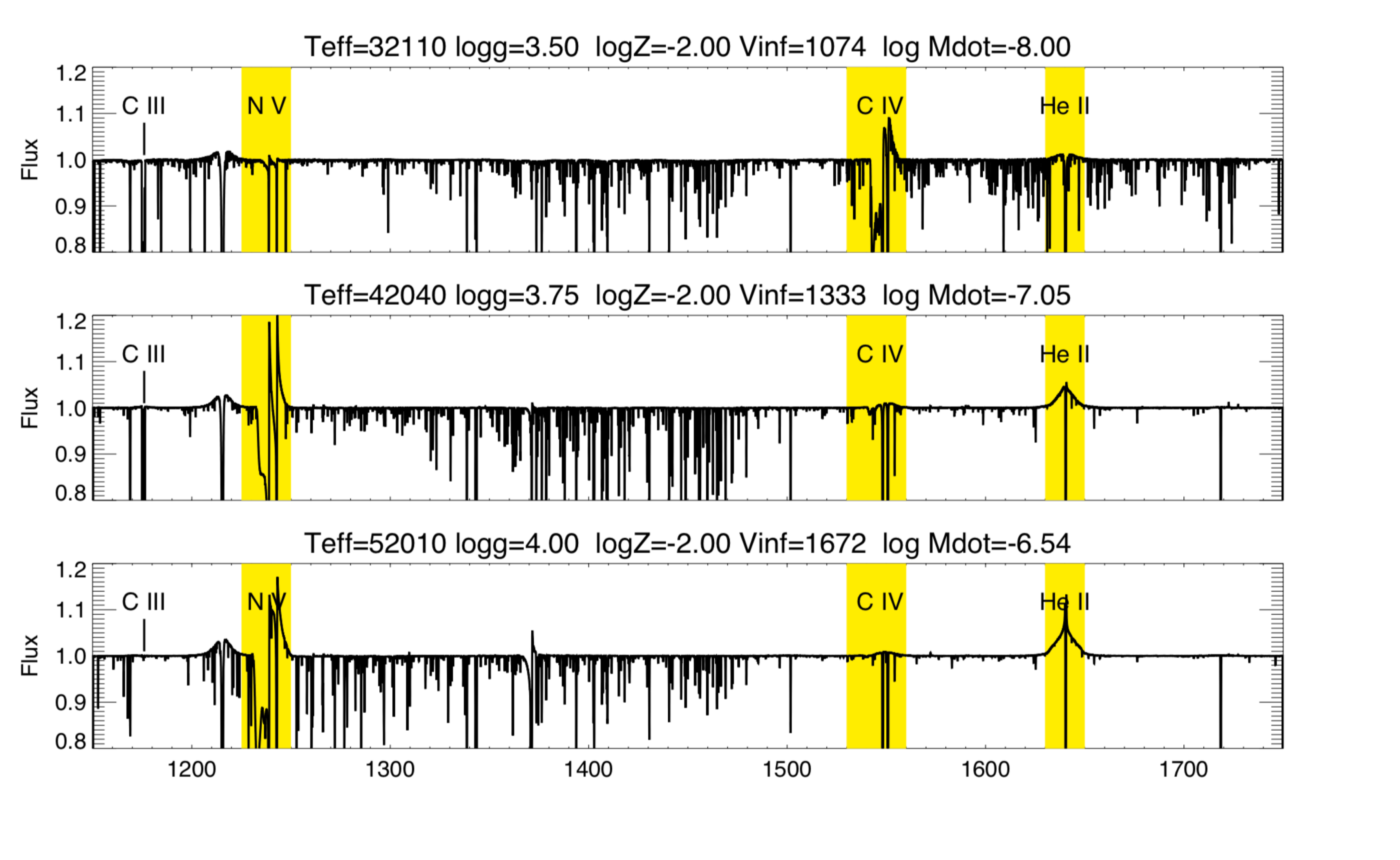}
\caption{CMFGEN models of the far-UV spectra of main-sequence O-type stars. The stellar parameters are given above each model. They are: effective temperature, log g, log Z,  terminal velocity of the wind in km/s, log mass-loss rate in solar masses per year. In a single star, the N~V doublet and     C~IV doublet cannot both be strong, whereas in a star cluster or galaxy, both features can be strong. As He II emission is formed in the stellar wind, He II $\lambda$1640 is a broad feature, easily distinguished from nebular emission lines which are narrow.}
\end{center}
\end{figure}

Far-UV spectra are similarly informative about the circumstellar,  interstellar, and circumgalactic medium of primitive galaxies. Figures 2-3 show the far-UV spectra of 9 galaxies having  nebular oxygen abundances, log O/H + 12 = 7.0 - 8.3.   Spectral regions encompassing the C IV $\lambda\lambda1548, 1550$ doublet, He~II $\lambda 1640$, 
and O~III] $\lambda\lambda 1661, 1666$ are highlighted in yellow.  The spectra are arranged in order of increasing oxygen abundance. In Figure 2, covering log Z = 7.0-7.8, not only nebular He~II $\lambda 1640$ (IP=54 eV) is in emission but also nebular C~IV doublet (IP=64 eV) in emission superimposed on the stellar spectrum, in which C~IV doublet has a P Cygni profile characteristic of a radiatively driven, high-velocity stellar wind. In Figure 3, covering log Z=7.8-8.2, nebular He~II $\lambda 1640$ emission weakens with increasing metallicity to the point that it is undetectable at log Z=8.2. At the same time, broad stellar emission, characteristic of Wolf-Rayet stars, becomes strong in two of the galaxies. 

\begin{figure}[h]
\begin{center}
\includegraphics[width=5.in]{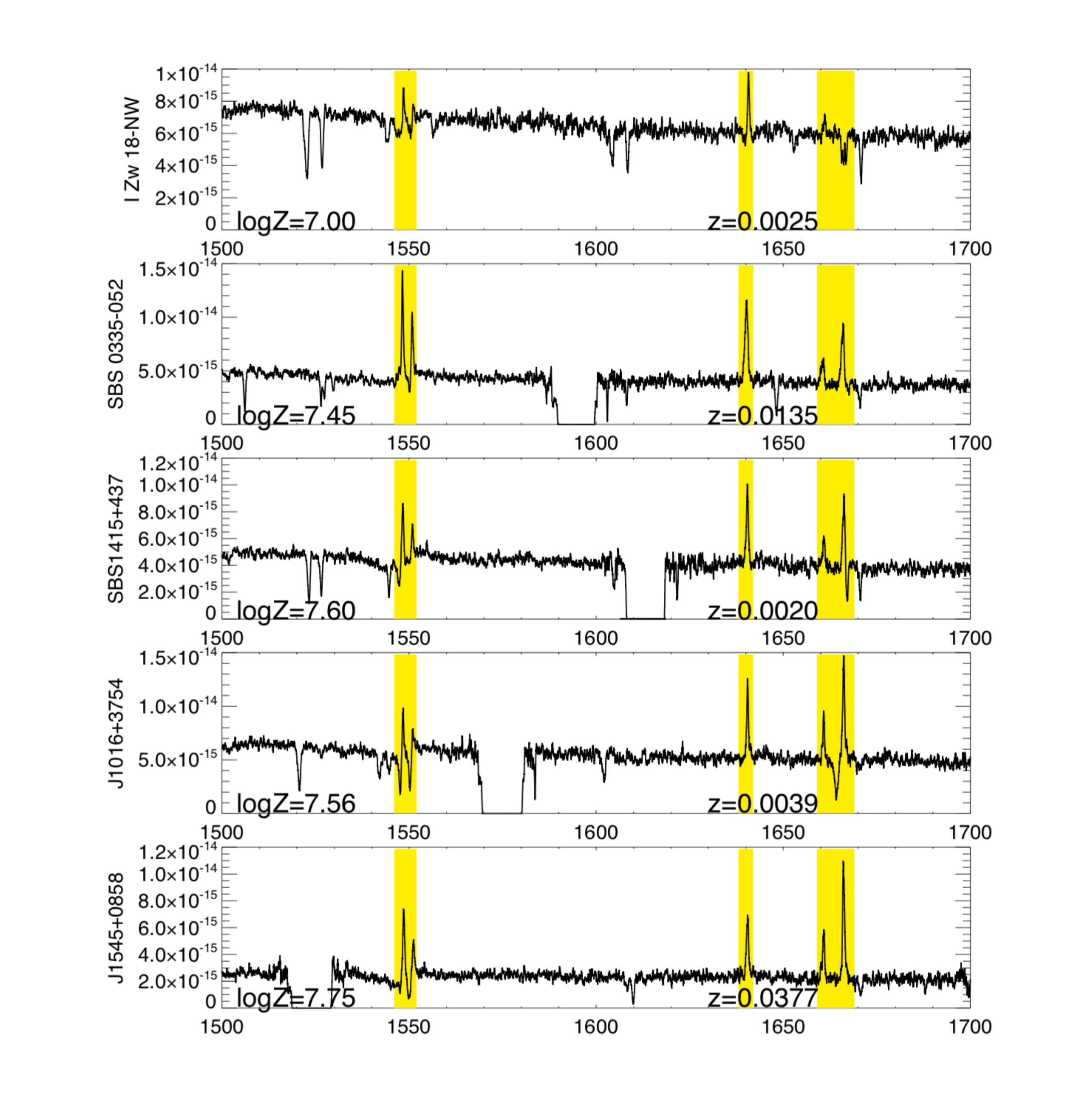}
\caption{Hubble/COS far-UV spectra of 9 low-Z galaxies galaxies. Spectral regions encompassing the C~IV $\lambda\lambda1548, 1550$ doublet, He~II $\lambda 1640$, and O~III] $\lambda\lambda 1661, 1666$ are highlighted. The galaxy ID is given on the Y-axis of the plot, and its redshift, at the bottom right of the plot. The nebular abundance of oxygen, log O/H+12, is listed at the bottom left, denoted as log Z. The spectra are arranged in order of increasing metal abundance. The spectra of \izwnw were obtained directly in programs ID=11523, 12028 (PI:Green), and most of the others in program ID=14120 (PI:Brinchmann). The two exceptions are SBS 0335-053 and SBS 1415+437, which were obtained from the Hubble archive (ID 15193, PI: Aloisi).}
\end{center}
\end{figure}

\pagebreak 

\begin{figure}[h]
\begin{center}
\includegraphics[width=5.in]{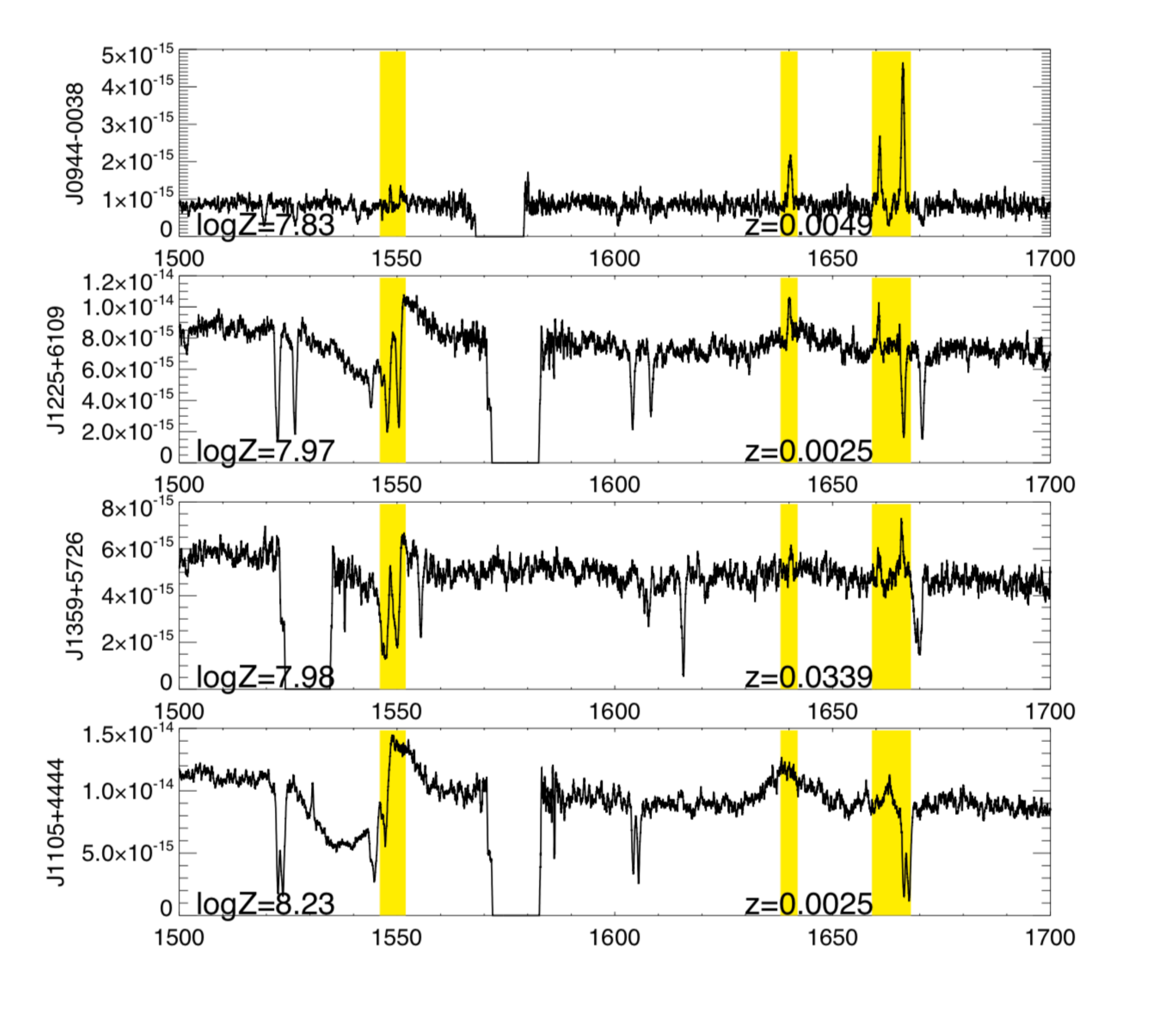}
\caption{Far-UV spectra of galaxies spanning log Z=7.83-8.23.}
\end{center}
\end{figure}

\section{Discussion and Conclusions}

\textit{How are we to understand this metallicity sequence of spectra?} To find out, we used CLOUDSPEC (Hubeny et al. 2000), a combination of CLOUDY (Ferland et al. 2017) and SYNSPEC (Hubeny \& Lanz, 1995)  to model the far-UV spectra of galaxies for 3 scenarios. In each scenario, there is a 5 Myr-old primitive stellar population (called ``SPop'' in Fig. 4) like \izwnw. In the first scenario (red), there is a dormant black-hole companion or no black hole at all. In the second (green), there is is a black-hole disk radiating X-rays at a rate of $10^{39}$ erg/s, and in the third (blue), a X-ray source ten times more luminous. Figure 4 shows the results. Nebular emission by He~II $\lambda 1640$ is not present when there is no black hole, it is detectable in the case of $L_x=10^{39}$ erg/s, and it quite strong in the $L_x=10^{40}$ erg/s case. The C~IV doublet is seen in absorption when there is no black hole, the absorption lines are partially filled in for the case of $L_x=10^{39}$ erg/s, and strongly in emission at $L_x=10^{40}$ erg/s. Although the statistics are limited, we conclude that the presence of 
C~IV emission is associated with galaxies whose metallicity,$\ log Z \leq 7.83$, while nebular He~II emission is still present in galaxies until $\log Z > 8.0$. 

\pagebreak

\begin{figure}[h]
\begin{center}
\includegraphics[width=4.in]{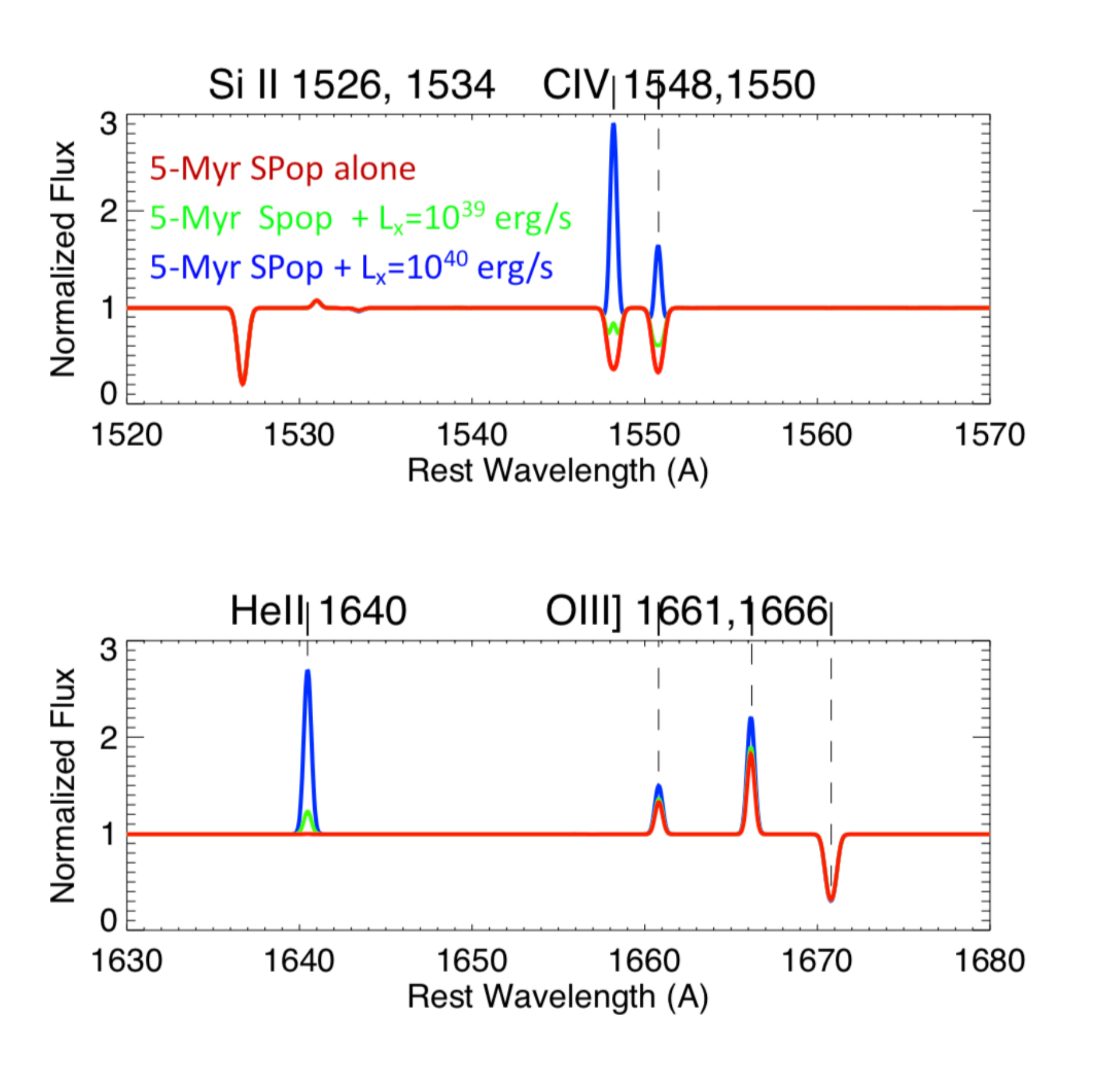}
\caption{Sample CLOUDSPEC model (Hubeny et al. 2000) of a primitive galaxy like \izwnw showing the effects of an embedded stellar black hole. Inputs to CLOUDSPEC include: full SED of a stellar cluster at age=5 Myr, log Z=7.2;  BH parameters: $L_x$ as shown, temperature of accretion disk, $T=10^6$ K; nebular parameters: election density, 
$N_e=10\, \rm{cm}^{-3}$, H I column density =$2\times 10^{21}$ cm${}^{-2}$, covering fraction=0.4, no dust.}
\end{center}
\end{figure}

\textit{Why should the presence of a black hole be correlated with metallicity?} One answer comes from a paper Heger et al. (2003), titled: ``How Massive Single Stars End Their Life''. It suggests that stars having a final mass greater than about 40 \Msun\ collapse directly into a black hole with little to no mass lost in an explosion. Since very low metallicity stars lose very little mass during their lifetime because of such  weak stellar winds, their final mass is 80\% of their initial mass (e.g. Groh et al. 2019), so stars formed with an initial mass of 50 \Msun\ or more will eventually collapse to a massive stellar black hole.

In this interpretation, we are viewing the former secondary orbiting a black hole, which formed from the collapsed primary, and the former secondary is now feeding the black hole via a black hole disk. The black hole disk is hot,
particularly at its inner edge where it is 1-2 million degrees, so its emission is centered on the X-ray spectral region. The reason why the disk is hot is because a stellar-mass black hole, as opposed to a supermassive black hole, forms a very deep potential well, so that accreted material is heated to such high temperatures.
 
\pagebreak

There are several COS far-UV spectra of primitive galaxies from the CLASSY program (Berg et al. 2022) in the Hubble archives that could be included. And combination with optical spectra from the SDSS and other ground-based telescopes would greatly benefit the statistics on primitive galaxies, especially if could account for primitive galaxies \textit{not} showing evidence of hard radiation as well as ones that do. There are also theoretical issues to be cleared up, mainly the evolution of massive stars in binary systems, since most hot, massive stars are in binary systems (Sana et al., 2012). Other factors such as the rotation and the C/O ratio should be  explored. 

We conclude that massive, extremely low metallicity galaxies are extensions from those of low metallicity, e.g. spectral lines are generally weaker, and stellar winds are weaker. However, primitive galaxies are qualitatively different from more metal-rich galaxies in showing evidence of hard radiation that can best be explained by the presence of one or more stellar black holes.

\end{document}